\documentclass[journal]{IEEEtran}

\usepackage{xcolor,soul,framed} 

\colorlet{shadecolor}{yellow}
\usepackage[pdftex]{graphicx}
\graphicspath{{../pdf/}{../jpeg/}}
\DeclareGraphicsExtensions{.pdf,.jpeg,.png}

\usepackage[cmex10]{amsmath}
\usepackage{array}
\usepackage{nccmath}
\usepackage{mdwmath}
\usepackage{mdwtab}
\usepackage{eqparbox}
\usepackage{url}
\usepackage{verbatim}
\usepackage{multirow}
\usepackage{multicol}
\usepackage{diagbox}
\usepackage[justification=centering]{caption}
\usepackage{algorithm}
\usepackage{algorithmic}
\usepackage{setspace}
\usepackage{amsfonts}
\usepackage{stfloats}
\usepackage{lipsum}
\usepackage{mathtools}
\usepackage{cuted}
\usepackage{pifont}
\usepackage{subfig}

\usepackage{caption}
\usepackage{amsthm}

\usepackage{fancyhdr}
\fancypagestyle{firstpage}{%
  \lhead{\textbf{This work has been submitted to the IEEE for possible publication.  Copyright may be transferred without notice, after which this version may no longer be accessible.}}
}

\newtheoremstyle{mystyle}
  {}
  {}
  {\upshape}
  {}
  {\bfseries}
  {:}
  { }
  {}

\theoremstyle{mystyle}

\newtheorem{myDef}{Definition}

\begin{document}
\bstctlcite{IEEEexample:BSTcontrol}
    \title{Multi-User Multi-Application Packet Scheduling for Application-Specific QoE Enhancement Based on Knowledge-Embedded DDPG in 6G RAN}

\author{Yongqin~Fu,~\IEEEmembership{Graduate Student Member,~IEEE,} and 
      Xianbin~Wang,~\IEEEmembership{Fellow,~IEEE}\\

\thanks{Y. Fu and X. Wang are with the Department of Electrical and Computer Engineering, Western University, London, ON N6A 5B9, Canada (e-mail: yfu335@uwo.ca; xianbin.wang@uwo.ca).}
\thanks{Corresponding author: Xianbin Wang.}
}
 


\maketitle

\thispagestyle{firstpage}

\begin{abstract}
The rapidly growing diversity of concurrent applications from both different users and same devices calls for application-specific Quality of Experience (QoE) enhancement of future wireless communications. Achieving this goal relies on application-specific packet scheduling, as it is vital for achieving tailored QoE enhancement by realizing the application-specific Quality of Service (QoS) requirements and for optimal perceived QoE values. However, the intertwining diversified QoE perception mechanisms,
fairness among concurrent applications, and the impact of network dynamics inevitably complicate tailored packet scheduling. To achieve concurrent application-specific QoE enhancement, the problem of multi-user multi-application packet scheduling in downlink 6G radio access network (RAN) is first formulated as a Markov decision process (MDP) problem in this paper. For solving this problem,  a deep deterministic policy gradient (DDPG)-based solution is proposed. However, due to the high dimensionalities of both the state and action spaces, the trained DDPG agents might generate decisions causing unnecessary resource waste. Hence, a knowledge embedding method is proposed to adjust the decisions of the DDPG agents according to human insights. Extensive experiments are conducted, which demonstrate the superiority of DDPG-based packet schedulers over baseline algorithms and the effectiveness of the proposed knowledge embedding technique.

\end{abstract}
\begin{IEEEkeywords}
Multi-user multi-application packet scheduling, application-specific QoE enhancement,  DDPG, knowledge embedding, 6G RAN
\end{IEEEkeywords}

\section{Introduction}
In recent years, a wide variety of new wireless-enabled applications have emerged, including both consumer-centric applications such as ultra high definition (UHD) video streaming \cite{salva20185g}, virtual reality \cite{10217163} as well as vertical industrial applications like robotic control \cite{ebert2018visual} and smart factory \cite{10505142}. These new applications and the traditional ones, including web browsing and voice over Internet Protocol (VoIP) \cite{6496406} have distinct QoE perception and evaluation schemes, which are related to application-specific quality of service (QoS) metrics, such as data transmission rate, downlink end-to-end latency and packet loss rate. In general, each wireless-enabled application running on a same user equipment (UE) has its unique combination of QoS requirements, reflecting their relative priorities. In particular, each wireless-enabled application also has its specific QoE perception mechanism and QoE-QoS relationship. Therefore, maximizing application-specific QoE provisioning by meeting its unique QoS requirement becomes the essential goal of resource management in 6G networks, which aims at optimizing resource management in order to maximize the perceived QoE values of all concurrent applications simultaneously. However, supporting these applications could cause dramatic variation of resource demands across different UEs. Due to the constraint of total available resource, maintaining fairness among different UEs and their running applications thus becomes a critical requirement of application-specific QoE enhancement.

Packet scheduling is an important resource management task, which has direct impact on QoE provisioning. In the downlinks radio access networks, packet scheduling determines the specific packet to be transmitted on each physical resource block (PRB) in the forthcoming transmission time interval (TTI). Apparently,  packet scheduling greatly influences applications' perceived QoS and QoE. Hence, packet scheduling is essential for ensuring fairnesses among UEs and their running applications. In order to perform packet scheduling, many factors need to be taken into consideration, including UEs' channel conditions and the status of each application's specific buffer, which often change over time. Therefore, making packet scheduling decisions timely can be challenging in order to successfully cope with the fast-changing environment. In addition, the diversified QoE perception mechanisms of applications unavoidably makes the problem of packet scheduling even more complicated to solve.

Hence, for maximizing concurrent application-specific QoE provisioning, the problem of multi-user multi-application packet scheduling is investigated in this paper based on the scenario of downlink transmission in 6G RAN. Since that different applications have different priorities, it's important to ensure both the fairness among the concurrent applications running on each UE and the fairness among different UEs, which are measured by intra-UE fairness and inter-UE fairness in this paper, respectively. Moreover, maximizing the sum of inter-UE fairness over one time period is selected as the optimization objective. This problem is first formulated as a sequential decision-making problem.

Evidently, the problem of multi-user multi-application packet scheduling in downlink RAN has two important properties. First, the packet scheduling policy taken by the packet scheduler during a specific time slot doesn't merely influence the immediately acquired reward but also influence the buffer statuses of applications in the subsequent time slot. Second, the inputs taken by the packet scheduler are determined not solely by the packet scheduling policies generated by the packet scheduler in the previous time slots, but also by the environment, which changes dynamically and randomly. For example, the packet arrival of applications and CQI values of UEs change randomly over time. Therefore, the original problem is further reformulated as a Markov decision process (MDP) \cite{puterman2014markov} problem, which could better capture the two important properties of the problem of multi-user multi-application packet scheduling in downlink 6G RAN.

Deep reinforcement learning (DRL) is a kind of techniques which is suitable to solve the MDP problems and has achieved great success in some fields such as gaming and robotic control. Once a DRL agent has been trained, it could generate packet scheduling policy very quickly, which makes it suitable to cope with the rapidly-changing environment and meet the stringent time requirement for packet scheduling.
However, despite their potentials, deep reinforcement learning techniques might not achieve satisfactory performance on this problem. The dimensionalities of both the state and action spaces of this problem become very high as the numbers of UEs and their applications grow. Due to the high dimensionalities of state and action spaces, it is difficult for DRL agents to grasp some human insights through exploration and exploitation by themselves within a limited number of training episodes, which inevitably restricts their performance.

Hence, in order to improve the performance of DRL agents, we come up with a novel idea, which is called knowledge embedding. Knowledge embedding means adjusting the actions generated by by DRL agents according to some human insights, such as no PRB should be allocated to one application which is not running or belongs to a user with the worst channel quality. Utilizing the proposed knowledge embedding method, resource waste could be greatly reduced and performance could be improved as well.

The main contributions of this paper could be summarized as follows.
\begin{enumerate}
    \item A novel concept of application-specific QoE enhancement in downlink 6G RAN is first conceptualized  while ensuring the fairness among different UEs with different applications. For achieving application-specific QoE enhancement in downlink 6G RAN, the problem of multi-user multi-application packet scheduling is formulated as a sequential decision-making problem with the objective of maximizing the sum of inter-UE fairnesses during a time period. In order to better capture the properties of multi-user multi-application packet scheduling, the original problem is further reformulated into a Markov decision process (MDP) problem.
    \item Due to the high dimensionalities of both the state and action spaces as well as the dynamically and randomly changing environment, the formulated MDP problem is difficult to solve. Hence, a deep deterministic policy gradient (DDPG)-based solution is designed for solving this problem, which could generate packet scheduling policy timely to cope with the fast-changing environment.
    \item  A novel knowledge embedding method is proposed for improving the performance of trained DDPG agents by adjusting the actions generated by them according to some human insights to avoid resource waste.
    \item In order to conduct performance evaluation more realistically, five representative applications are considered, and parametric QoE estimation method is utilized to estimate the perceived QoE values based on some common QoS indicators. For fairly comparing the performance of the packet scheduling algorithms, a test dataset is created which is consisted of 100 episodes with time length of 1 minute. Extensive experiments are conducted, whose results confirm the superiority of DDPG-based packet schedulers over the baseline algorithms. Moreover, the simulation results confirm the effectiveness of knowledge embedding on improving the performance of DDPG-based packet schedulers.
\end{enumerate}

\section{Related Work}
\subsection{Conventional Downlink Packet Scheduling Techniques in RAN}
According to \cite{6226795}, the conventional packet scheduling techniques for downlink transmission in RAN could be classified into five categories, which are channel-unaware \cite{stolyar2001largest}, channel-aware but QoS-unaware \cite{1543653}, channel-aware and QoS-aware \cite{5956352}, semi-persistent (for voice over Internet protocol (VoIP) \cite{5206336} and energy-aware \cite{5962571}.

\subsection{Downlink Packet Scheduling in RAN Based on Deep Reinforcement Learning}
The existing downlink packet scheduling techniques based on deep reinforcement learning could be roughly classified into two categories, which are indirect and direct DRL packet scheduling techniques. 

The indirect DRL packet scheduling techniques utilize DRL to facilitate the selection of parameters or conventional packet scheduling techniques to generate packet scheduling policy using non-DRL approaches. In other words, the indirect DRL packet scheduling techniques don't utilize DRL to allocate PRBs to UEs / applications directly. In \cite{8927868}, the authors consider the scenario of single-carrier packet scheduling in a single cell network. Moreover, three DRL techniques are proposed for selecting one UE to be scheduled on a resource block group (RBG) during each TTI, which are direct-, dual- and expert-learning. The authors in \cite{9427224} propose a knowledge-assisted DRL algorithm for downlink packet scheduling of time-sensitive traffic in 5G NR. In addition, the actor network of the proposed algorithm is trained to output the set of UEs to be scheduled. In \cite{9851620}, the problem of delay-oriented packet scheduling in multi-cell downlink 5G new radio (NR) networks is investigated. Moreover, a DRL-based packet scheduling framework is proposed, where the actor network is trained to output the scheduling priority of each active traffic flow. Aiming at minimizing packet delays and packet drop rates, the authors in \cite{8425580} propose a novel downlink packet scheduling framework which could select suitable scheduling rules according to time-varying states, utilizing reinforcement learning principles.
 
Different from indirect DRL packet scheduling techniques, direct DRL packet scheduling techniques utilize DRL to learn and generate packet scheduling policy directly. In other words, the direct DRL packet scheduling techniques utilize DRL to allocate PRBs to UEs / applications directly. In \cite{9217110}, the authors propose a DDPG-based solution for downlink packet scheduling considering UEs with delay-sensitive applications, with the objective of minimizing UEs' experienced queuing delay. The authors in \cite{9838349} design a DDPG-based downlink packet scheduling algorithm considering the scenario where a priority UE coexists with normal UEs. In addition, the actor network of DDPG algorithm is trained to output the percentages of resource blocks to be allocated to the UEs. In \cite{9099267}, the authors consider the coexistence of eMBB and uRLLC services and propose a DDPG-based solution for jointly optimizing the allocation of resource blocks and puncturing positions of uRLLC traffic, to achieve tradeoff between UEs' experienced QoS values of the two types of services. The authors in \cite{9209313} train a deep pointer network to learn to generate the scheduling decision on one resource block at a time with the objective of maximizing throughput and Jain's fairness index (JFI), utilizing the advantage actor-critic (A2C) approach. In \cite{9652950}, a pointer network is trained to generate the scheduling decisions on all the resource blocks simultaneously, utilizing the DDPG approach without exploration noise. The authors in \cite{9110842} design a double deep Q network (DDQN)-based solution, which generates a scheduling decision for one RBG at a time. In \cite{9814547}, a deep Q-network (DQN)-based downlink scheduler is proposed for 5G NR with the objective of maximizing throughput and fairness among UEs, which allocates one RBG at a time until all the RBGs are allocated.  The authors in \cite{9322560} propose a DQN-based packet scheduling method utilizing a recurrent neural network (RNN) for generating scheduling decisions according to observed states, aiming at minimizing the average packet delay. In \cite{9120729}, an A2C-based solution is designed to schedule UEs on one RBG at a time, aiming at optimizing throughput, fairness among UEs and packet loss rate simultaneously. The authors in \cite{10125013} propose a DDQN-based approach for downlink packet scheduling with guarantee of per-packet delay, aiming at optimizing the average queue length and the average output gain concurrently. In \cite{10431724}, the authors investigate the problem of multi-user latency-constrained scheduling considering resource constraints. In order to deal with partial observability and scalability, a novel data-driven DRL method is proposed, utilizing Lagrangian dual approach and delay-sensitive queues to cope with resource and latency constraints, respectively.

\section{System Model}
\begin{figure*}[hbtp]
  \begin{center} 
  \includegraphics[width=5.5in]{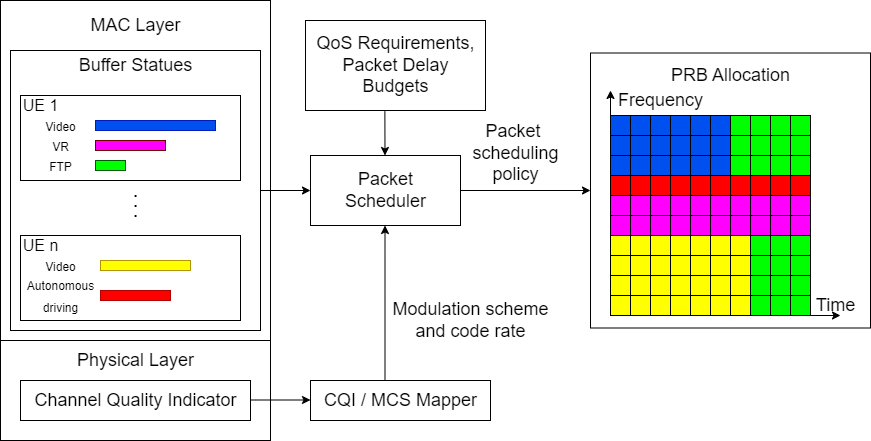}\\
  \caption{Illustration of the process of multi-user multi-application packet scheduling in downlink 6G RAN. The packet scheduler takes information from both physical and MAC layers, as well as the QoS requirements and packet delay budgets as inputs to generate packet scheduling policy. The CQI values of UEs are first transformed into modulation schemes and code rates by CQI / MCS mapper, then forwarded to packet scheduler as inputs.}\label{PacketScheduling}
  \end{center}
\end{figure*}

In this paper, we consider the scenario of downlink transmission in a single-cell multi-user RAN. The process of multi-user multi-application packet scheduling in downlink 6G RAN is illustrated in Fig. \ref{PacketScheduling}. As illustrated in Fig. \ref{PacketScheduling}, the information of both physical and medium access control (MAC) layers are utilized for decision making. For physical layer information, channel quality indicators (CQIs) are considered. CQIs are reported by UEs periodically indicating their channel conditions, ranging from 0 to 15. The set of CQI values is denoted by $I$. Specifically, the CQIs are taken as the inputs of CQI / Modulation and Coding Scheme (MCS) mapper.

\begin{myDef}
\textbf{CQI/MCS mapping}

The CQI/MCS mapping is defined as two one-to-one mappings, $\mu: I \rightarrow M$ and $\nu: I \rightarrow C$, where $M$ denotes the set of modulation schemes and $C$ denotes the set of code rates.
\end{myDef}

In addition, for a specific modulation scheme $m \in M$, the number of bits transmitted by one symbol is denoted by $W_m$. The modulation schemes and code rates of UEs are utilized as inputs for the packet scheduler to make policy. 

For MAC layer information, the buffer statuses of applications of UEs are considered, such as buffer length and age of head-of-line packet. In addition, the QoS requirements of applications are also utilized as inputs for packet scheduler to make policies. 

We assume that downlink packet scheduling is performed in discrete time slots, whose length is equal to one TTI, denoted as $\Delta T$. A time period  is consisted of $T$ time slots, whose set is denoted as $\mathcal{T} = \{1, \dots, T\}$. We assume that there exist $U$ UEs within the cell, whose set is denoted by $\mathcal{U} = \{1, \dots, U\}$. Let $\phi_u(t)$ denote the variable indicating whether UE $u$ is active during time slot $t$:

\begin{equation}
    \phi_u(t) = \left \{
    \begin{array}{lr}
    1,\quad \textrm{if the $u$-th UE is active}\\
    \quad \quad \textrm{during time slot $t$} \\
    0, \quad \textrm{otherwise}
    \end{array}
    \right . 
\end{equation}

The set of active UEs during time slot $t$ is denoted as $\mathcal{U}(t)$. Hence,  $\mathcal{U}(t) = \{u | \phi_u(t) = 1\}$. The CQI value of UE $u$ during time slot $t$ is denoted as $I_u(t)$. Moreover, we assume that each UE carries up to $K$ applications, denoted by $\mathcal{A} = \{1, 2, \dots, K\}$. We assume that different applications have different priority weights and the higher priority weight indicates the higher priority. The priority weight of application $k$ is denoted by $P_k$. Let $\sigma_u^k(t)$ denote the variable indicating whether the application $k$ of UE $u$ is active during time slot $t$ as follows:

\begin{equation}
    \sigma_u^k(t) = \left \{
    \begin{array}{lr}
    1,\quad \textrm{if the $k$-th application of the $u$-th UE}\\
   \quad \quad \textrm{is active during time slot $t$} \\
    0, \quad \textrm{otherwise}
    \end{array}
    \right . 
\end{equation}

We denote the set of active applications of UE $u$ during time slot $t$ as $\mathcal{A}_u(t)$. Hence, $\mathcal{A}_u(t) = \{k | \sigma_u^k(t) = 1\}$. 
We assume that the frequency band of the base station is divided into $B$ PRBs, whose set is denoted by $\mathcal{B} = \{1, \cdots, B\}$. The allocation policy on PRB $b$ and the application $k$ of UE $u$ during time slot $t$ is denoted as:
\begin{equation}
    \rho_b^{u,k}(t) = \left \{
    \begin{array}{lr}
    1,\quad \textrm{if the $b$-th PRB is allocated to }\\
    \quad \quad \textrm{application $k$ of UE $u$ during}\\
    \quad \quad \textrm{time slot $t$} \\
    0, \quad \textrm{otherwise}
    \end{array}
    \right . 
\end{equation}

The number of useful bits transmitted by a PRB allocated to UE $u$ during time slot $t$ is calculated as  
\begin{equation}
V_u(t) = W_{\mu(I_u(t))}\nu(I_u(t)) \times 12 \times 14, 
\end{equation}
where $I_u(t)$ indicates the CQI value of UE $u$ during time slot $t$ and $W_i$ denotes the number of bits per symbol carries for modulation scheme $i$, and 12 and 14 are the amounts of subcarriers and orthogonal frequency division multiplexing (OFDM) symbols per PRB, respectively.
The number of PRBs allocated to application $k$ running on UE $u$ during time slot $t$ is denoted by $N_u^k(t)$. Therefore, 
\begin{equation}
N_u^k(t) = \sum_{b = 1}^{B}\rho_b^{u,k}(t). 
\end{equation}
Hence, the data transmission rate experienced by application $k$ running on UE $u$ during time slot $t$ is 
\begin{equation}
R_u^k(t) = \frac{N_u^k(t)V_u(t)}{1000} \; \textrm{Mbps}. 
\end{equation}
We assume that the packet size of application $k$ is denoted by $s_k$ bits and the packet delay budget of application $k$ is $D_k$ TTIs, beyond which a packet will be regarded as obsolete and discarded. The queuing latency experienced by application $k$ running on UE $u$ during time slot $t$ is defined as 
\begin{equation}
  L_{u,k}^{queu}(t) = t - F_u^k(t),  
\end{equation}
where $F_u^k(t)$ denotes the number of 'finished' slots of application $k$ running on UE $u$ during time slot $t$. Specifically, time slot $t_1$ is regarded as 'finished' in time slot $t$ if all the packets arrived in time slot $t_1$ have been transmitted or discarded during time slot $t$. The transmission latency experienced by application $k$ running on UE $u$ during time slot $t$ is
\begin{equation}
L_{u,k}^{tran}(t) = \frac{S_k}{N_u^k(t)V_u(t)} \; \textrm{ms}. 
\end{equation}
Moreover, we assume that the average latency of application $k$ from server to the base station is $L_k^{sb}$. The latency caused by baseband signal processing at the BS and the propagation latency are neglected as they are not dominating components. Hence, the total downlink end-to-end latency experienced by application $k$ running on UE $u$ during time slot $t$ is 
\begin{equation}
L_u^k(t) = L_k^{sb} + L_{u,k}^{queu}(t) + L_{u,k}^{tran}(t).
\end{equation}

We assume that the packet arrival of application $k$ follows Poisson process and the mean packet arrival rate is $\lambda_k$ packets per TTI. The number of packets of application $k$ running on UE $u$ arrived during time slot $t$ is denoted by $X_u^k(t)$. The number of packets of application $k$ running on UE $u$  which were discarded during time slot $t$ is denoted by $Y_u^k(t)$. In addition, we assume that the length of time window for measurement of packet loss rate is $T_p$ TTIs. Hence, the packet loss rate experienced by application $k$ of UE $u$ during time slot $t_o$ is defined as:
\begin{equation}
    H_u^k(t_o) = \frac{\sum_{t=\textrm{max}(t_o - T_p,\; 0)}^{t=t_o}Y_u^k(t)}{\sum_{t=\textrm{max}(t_o - T_p,\; 0)}^{t=t_o}X_u^k(t - D_k)}.
\end{equation}

We assume that each application $k \in \mathcal{A}$ has its individual QoS requirements on data transmission rate, downlink end-to-end latency and packet loss rate, denoted by $R_k^{req}$, $L_k^{req}$ and $H_k^{req}$, respectively. The QoE value experienced by application $k$ of UE $u$ during time slot $t$ is denoted as $Q_u^k(t)$.

\begin{myDef}
\textbf{Intra-UE fairness}

We define the intra-UE fairness for packet scheduling of UE $u$ during time slot $t$ as the weighted average of active applications' QoE values multiplied by QoS requirement satisfaction indicators:
\begin{equation}
    f_u(t) = \frac{\sum_{k \in \mathcal{A}_u(t)}\beta_u^k(t)I_kQ_u^k(t)}{\sum_{k \in \mathcal{A}_u(t)}I_k},
\end{equation}
where $\beta_u^k(t)$ indicates whether the QoS requirements are satisfied during time slot $t$:
\begin{equation}
    \beta_u^k(t) = \left \{
    \begin{array}{lr}
    1,\quad \textrm{if} \; R_u^k(t) \ge R_k^{req}\;\textrm{and} \:  L_u^k(t) \le L_k^{req} \; \\
   \quad \quad \textrm{and}\; H_u^k(t) \le H_k^{req}\\
    0, \quad \textrm{otherwise}\\
    \end{array}
    \right . 
\end{equation}
\end{myDef}

\begin{myDef}
\textbf{Inter-UE fairness}

We define the inter-UE fairness for multi-user multi-application packet scheduling during time slot $t$ as the average of all the active UEs' intra-UE fairnesses:
\begin{equation}
    f(t) =\frac{\sum_{u \in \mathcal{U}(t)}f_u(t)}{|\mathcal{U}(t)|},
\end{equation}
where $|Z|$ means the number of elements in set $Z$.
\end{myDef}

\section{QoE Modelling}
Without loss of generality, we select five representative applications, which are File Transfer Protocol (FTP) \cite{8975755}, ultra high definition (UHD) video streaming, web browsing, online gaming and voice over Internet protocol (VoIP), whose indexes are 1 to 5, respectively. Since QoE is influenced by many factors, it is not easy to calculate the perceived QoE values. In this paper, we utilize parametric QoE estimation method to estimate the perceived QoE values based on some common QoS indicators. Next, we will introduce the QoE modelling of the five applications.

\subsection{QoE Modelling of FTP}
For FTP application, we utilize the following formula introduced in \cite{tsolkas2017survey}:

\begin{equation}
    Q_u^1(t) = \left \{
    \begin{array}{lr}
    1,\quad \textrm{if} \; R_u^1(t) < 8 \; \textrm{kbps}\\
    2.5037\textrm{log}_{10}(0.3136R_u^1(t)),\\
    \quad \quad \textrm{if} \; 8 \; \textrm{kbps} \le R_u^1(t) < 315 \; \textrm{kbps}\\
    5, \quad \textrm{if} \; 315 \; \textrm{kbps} \le R_u^1(t) 
    \end{array}
    \right . 
\end{equation}

\subsection{QoE Modelling of UHD Video Streaming}
For UHD video streaming application, we utilize the following formula introduced in \cite{nightingale20185g}:

\begin{equation}
    Q_u^2(t) = -0.891 + \frac{5.082}{\sqrt{\frac{M}{R_u^2(t)}}},
\end{equation}
where $M$ represents the maximum bandwidth requirement of UHD video streaming application, and we set $M = 15 Mbps$.

\subsection{QoE Modelling of Web Browsing}
For web browsing application, we utilize the following formula introduced in \cite{sousa2020survey}:

\begin{equation}
    Q_u^3(t) = 5 - \frac{578}{1 + (11.77 + 22.61\frac{R_u^3(t)}{S_w})^2},
\end{equation}
where $S_w$ represents the size of a web page, and we set $S_w = 3MB$.

\subsection{QoE Modelling of Online Gaming}
For online gaming application, we utilize the following formula introduced in \cite{slivar2014empirical}:

\begin{equation}
    Q_u^4(t) = 4.7059 - 0.00094L_u^4(t) - 5.83444H_u^4(t)
\end{equation}

\subsection{QoE Modelling of VoIP}

For QoE modelling of VoIP application, we first utilize the following formula introduced in \cite{5929300} to calculate the rating factor $RF$ of VoIP application of UE $u$ during time slot $t$.
\begin{equation}
    RF_u^5(t) = 93.355 - 95\frac{H_u^5(t)}{\frac{H_u^5(t)}{BurstR_u^5(t)} + 4.3},
\end{equation}
where $BurstR_u^5(t)$ represents the burst ratio of UE $u$ during time slot $t$, calculated using $BurstR_u^5(t) = G_u^5(t)(1 - H_u^5(t))$,
where $G_u^5(t)$ represents the average observed length of series of lost packets. Moreover, we assume that the length of time window for measuring burst ratio is $T_b$ TTIs. After calculating the rating factor, we utilize the following formula introduced in \cite{tsolkas2017survey} for QoE modelling:
\begin{equation}
    Q_u^5(t) = \left \{
    \begin{array}{lr}
    1,\quad \textrm{if} \; RF_u^5(t) < 0\\
    1 + 0.035RF_u^5(t) + RF_u^5(t)(RF_u^5(t)\\
    - 60) (100 - RF_u^5(t))\times 7 \times 10^{-6}\\
    \quad \quad \textrm{if} \; 0 \le RF_u^5(t) \le 100\\
    4.5, \;\textrm{if} \; RF_u^5(t) > 100
    \end{array}
    \right . 
\end{equation}

\section{Problem Formulation}
The problem of multi-user multi-application packet scheduling in downlink 6G RAN is formulated as shown in (20). In (20a), we can find that the optimization goal is to maximize the sum of inter-UE fairnesses in the total $T$ time slots. Moreover, constraint (20b) constrain that each PRB can either be allocated or not allocated to each running application of each active UE. In addition, constraint (20c) constrain that each PRB can be allocated to at most one running application of an active UE during each time slot.

\begin{subequations}
\begin{flalign}
& \quad \quad \quad \quad \quad \quad \quad \quad \textrm{(P0)} \; \textrm{max} \; \sum_{t = 1}^{T}f(t)\\
   & \textrm{subject}  \textrm{ to:}  \notag\\
    &\rho_b^{u,k}(t) \in \{0, 1\}, \forall u \in \mathcal{U}(t), \forall  k \in \mathcal{A}_u(t), \forall b \in \mathcal{B}, \forall t \in \mathcal{T}\\
    & \sum_{u \in \mathcal{U}(t)}\sum_{k \in \mathcal{A}_u(t)}\rho_b^{u,k}(t) \le 1, \notag \\ & \forall u \in \mathcal{U}(t),  \forall  k \in \mathcal{A}_u(t), \forall b \in \mathcal{B}, \forall t \in \mathcal{T}
\end{flalign}
\end{subequations}

\section{Problem Reformulation as a Markov Decision Process Problem}

In (20), the optimization objective is to not to maximize the immediate reward of each time slot but to maximize the expected cumulative reward. Hence, this problem is in fact a sequential decision-making problem \cite{kochenderfer2022algorithms}, where DRL has achieved great success. Hence, we choose to use DRL techniques to solve the formulated multi-user multi-application packet scheduling problem. Due to the lack of knowledge of the transition probability distribution and the high dimensionality of the action space, policy-based reinforcement learning techniques are suitable for solving the formulated problem.

In order to apply policy-based reinforcement learning techniques to solve the formulated problem, we first reformulate the problem as a Markov decision process (MDP) problem, represented as a 5-tuple $(S, A, T_a, R_a, \gamma)$. Specifically, $S$ means the state space, $A$ means the action space, $T_a$ means transition function, $R_a$ means reward function and $\gamma$ is the discount factor. The details are as shown below.

\subsection{State Space}
We abstract the state of application $k$ of UE $u$ during time slot $t$ into $(\mu_u^k(t), \kappa_u^k(t), \sigma_u^k(t), I_u(t))$, where $\mu_u^k(t)$ means buffer length of application $k$ of UE $u$ during time slot $t$, $\kappa_u^k(t)$ means age of head-of-line packet of application $k$ of UE $u$ during time slot $t$. Hence, the state at time slot $t$ is a $U \times K \times 4$ matrix, represented by $S(t)$.

\subsection{Action Space}
The action taken at time slot $t$ is a $U \times K$ matrix, represented by $A(t)$, where the element $A_{i,j}(t)$ indicates the number of PRBs allocated to the $j$-th application of the $i$-th UE during time slot $t$.

\subsection{Transition Function}
The transition function $T_a(s, s')$ describes the probability that the state changes from $s$ to $s'$ after action $a$ is selected. However, we have no prior knowledge of $T_a(s, s')$ and the DRL agent needs to learn it through interaction with the environment.

\subsection{Reward Function}
We select the inter-UE fairness function $f(t)$ as the reward function, which describes the immediate reward received in time slot $t$, which is determined by state $S(t)$ and action $A(t)$.

\subsection{Discount Factor}
Except for the overall optimization goal, a specific goal is required for guiding the action selection at each time slot, which is formulated using a discount factor $\gamma$, ranging from 0 to 1.

\subsection{Goal at Each Time Slot}
At time slot $t_0$, we set the goal as maximizing the expected cumulative reward $F(t_0)$, which is formulated as 
\begin{equation}
F(t_0) = \sum_{t=0}^{+\infty}\gamma^tf(t_0 + t).
\end{equation}

\section{Implementation of DDPG-Based Packet Scheduler}
In this paper, a policy-based deep reinforcemnet learning algorithm is implemented, which is deep deterministic policy gradient (DDPG). The details of the implementation of DDPG algorithm will be introduced in the following subsesctions.

\subsection{Design of Actor Network}
\begin{figure*}[hbtp]
  \begin{center} 
  \includegraphics[width=6.0in]{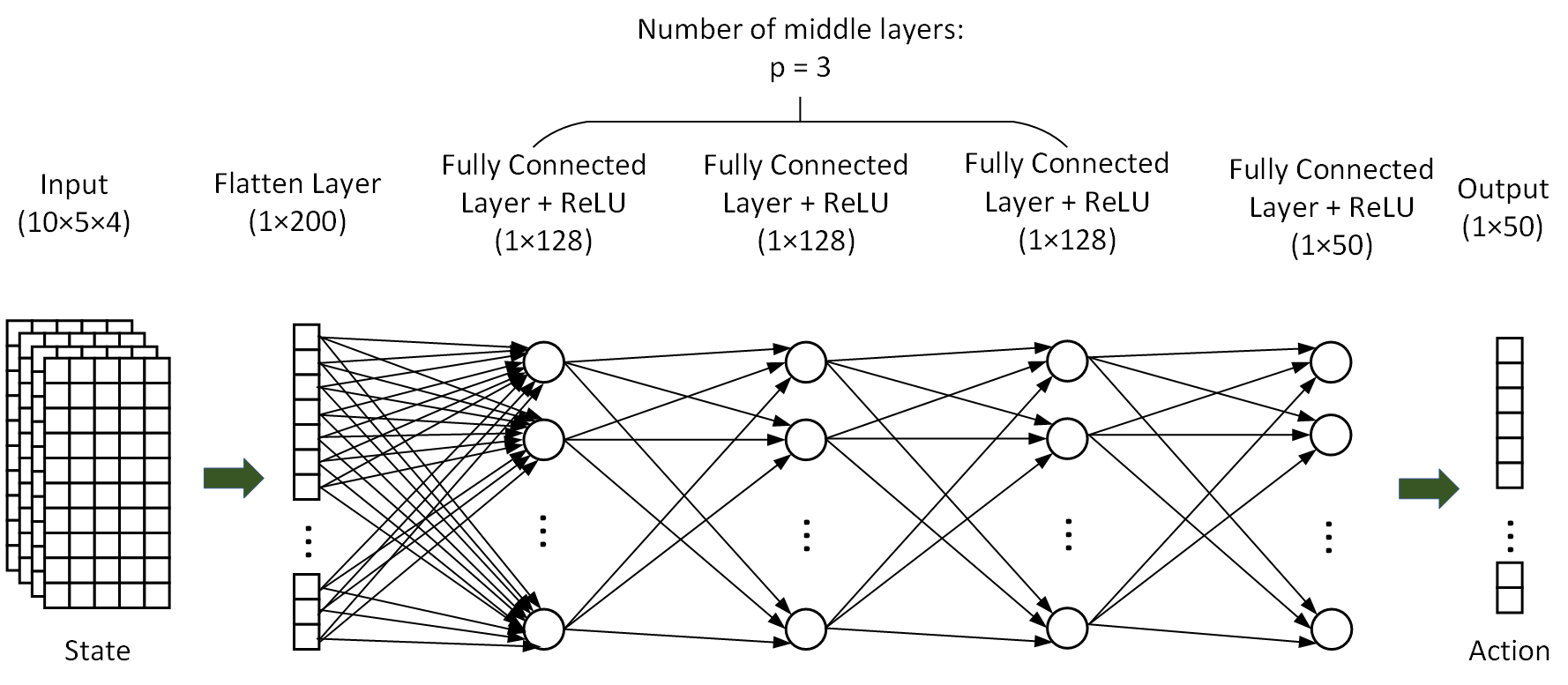}\\
  \caption{Illustration of the design of actor network of DDPG algorithm. The actor network is composed of one input layer, one flatten layer, a number of fully connected layers activated by ReLU function as middle layers (the number of middle layers is set to be 3 in this figure) , and one fully connected layer activated by ReLU function as output layer.}\label{actorNetwork}
  \end{center}
\end{figure*}

In DDPG algorithm, an actor network is utilized to generate an action according to the current state. The design of the actor network is illustrated in Fig. \ref{actorNetwork}.

In this paper, we set the total number of UEs to be 10 and the total number of applications to be 5. Hence, the state is a $10 \times 5 \times 4$ matrix, which is the input of the actor network. The flatten layer transforms the input matrix into a one-dimensional matrix, which is followed by four fully connected layers to generate the output, whose size is $1 \times 50$. Then the action is generated according to the output matrix, as shown in Alg. \ref{actionGeneration}, where round(x) means finding the integer closest to x.

\begin{figure*}[hbtp]
  \begin{center} 
  \includegraphics[width=6in]{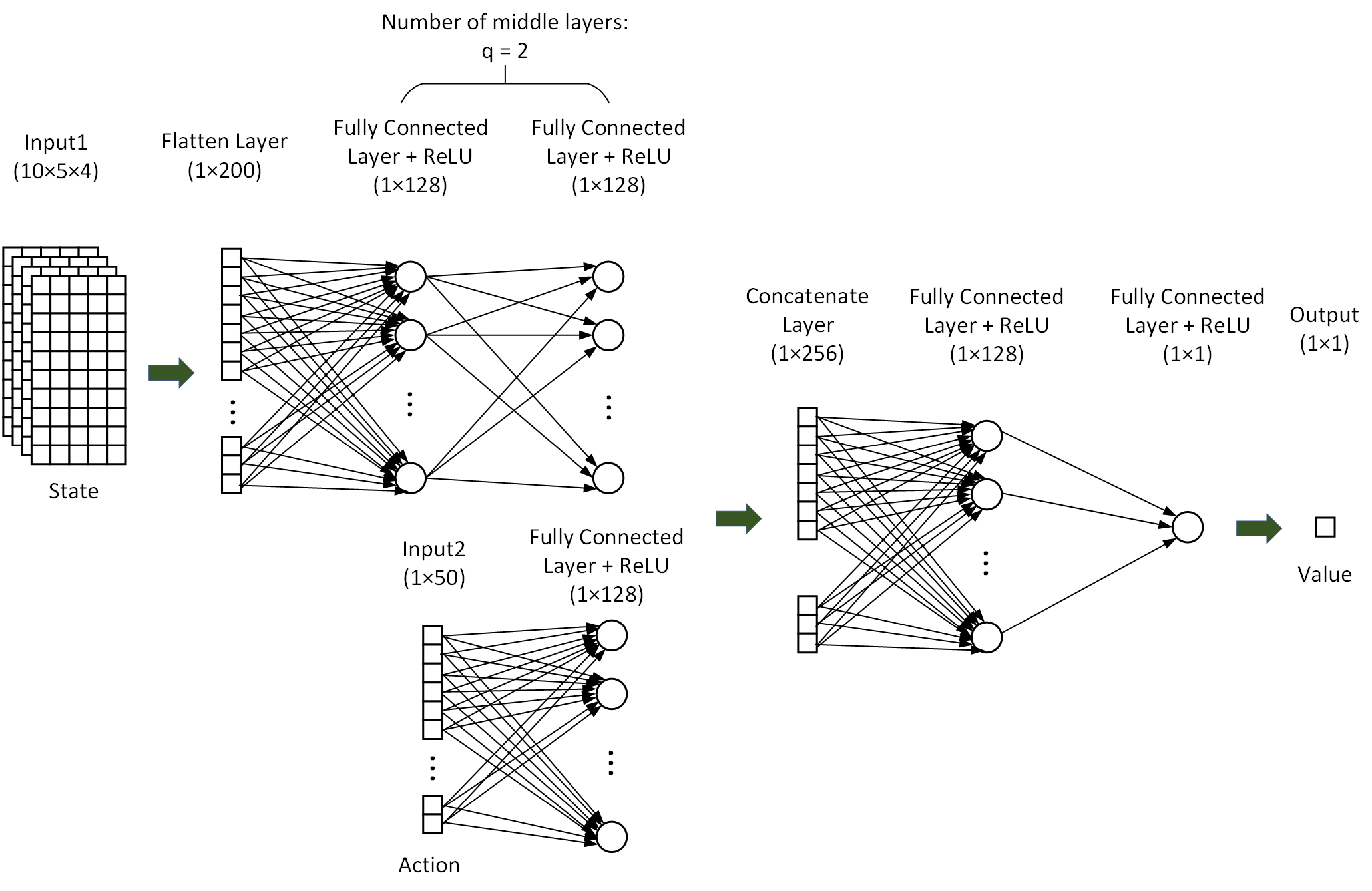}\\
  \caption{Illustration of the design of critic network of DDPG algorithm. The critic network contains two branches as it takes two separate inputs. The first branch is composed of one input layer, one flatter layer and a number of fully connected layers activated by ReLU function as middle layers (the number of middle layers is set to be 2 in this figure). The second branch is composed of one input layer and one fully connected layer activated by ReLU function. The two branches are connected to one concatenate layer followed by two fully connected layers activated by ReLU function to generate the output value of the input (state, action) pair.}\label{criticNetwork-DDPG}
  \end{center}
\end{figure*}

\begin{algorithm}[htbp]
\setstretch{1.35} 
\caption{Action Generation Algorithm}
\label{actionGeneration}
\begin{algorithmic}
\REQUIRE The output of actor network during time slot $t$, $O(t)$;\\
\STATE Initialize action at time slot $t$ as a $U \times K$ zero matrix, $A(t)$;\\
\STATE sum = 0;\\
\FOR{$i = 1$ to  $U$}
\FOR{$j = 1$ to $K$}
\STATE sum = sum + $|O_{(i-1) \times K + j}(t)|$;\\
\ENDFOR
\ENDFOR
\STATE remainder = $B$;\\
\FOR{$i = 1$ to $U$}
\FOR{$j = 1$ to $K$}
\IF{$i == U$ and $j == K$}
\STATE $A_{i,j}(t)$ = remainder;\\
\ELSIF{remainder == 0}
\STATE $A_{i,j}(t)$ = 0;\\
\ELSIF{sum == 0}
\STATE $A_{i,j}(t)$ = remainder;\\
\STATE remainder = 0;\\
\ELSE
\STATE $A_{i,j}(t)$ = round(remainder $\times \; |O_{(i-1) \times K + j}(t)|$ / sum);\\
\STATE remainder = remainder - $A_{i,j}(t)$;\\
\STATE sum = sum - $|O_{(i-1) \times K + j}(t)|$;\\
\ENDIF
\ENDFOR
\ENDFOR
\ENSURE Action at time slot $t$, $A(t)$;\\
\end{algorithmic}
\end{algorithm}

\subsection{Design of Critic Network}
 The design of critic network is illustrated in Fig. \ref{criticNetwork-DDPG}. We can find that the critic network takes both the state and action as inputs. Hence, it generates a value of (state, action) pair.
 
 \subsection{DDPG Algorithm}
 DDPG algorithm is shown in Alg. \ref{DDPG}.
 In Alg. \ref{DDPG}, four different neural networks are utilized, which are actor network, critic network, target actor network and target critic network.
 The parameters of actor, critic, target actor and target critic networks are represented by $\theta_a$, $\theta_c$, $\theta_a'$ and $\theta_c'$, respectively.
 
 Moreover, experience replay technique \cite{lin1992reinforcement} is utilized to increase the learning stability. During each time slot $t \in \{1, \cdots, T\}$, a minibatch of $m$ transitions are randomly sampled from the replay buffer, denoted as $\{S_i, A_i, f_i, S_{i+1} | \; i \in \{1, \cdots, m\}\}$. Then the loss of critic network is calculated as:

\begin{equation*}
  \mathcal{L}_c = \frac{1}{m}\sum_{i = 1}^{m}(f_i + \gamma \chi(S_{i+1}, O(S_{i+1}, \theta_a'), \theta_c') - \chi(S_i, A_i, \theta_c))^2,
\end{equation*}
where $\chi(S, A, \theta_c)$ denotes the output of the critic network with parameter $\theta_c$ and input state-action pair $(S,A)$. Moreover, $O(S, \theta_a)$ denotes the output of the action network with parameters $\theta_a$ and input $S$. Moreover, the parameters of the critic network are updated by minimizing $\mathcal{L}_c$.

The derivative of the actor network is calculated as 
\begin{equation}
d\theta_a \leftarrow \Delta_{\theta_a}\frac{1}{m}\sum_{i = 1}^m\chi(S_i, O(S_i, \theta_a), \theta_c).
\end{equation}
And the parameters of the actor network is updated using:
\begin{equation}
\theta_a \leftarrow \theta_a + \lambda(ep) \times d\theta_a.
\end{equation}

After updating the parameters of the actor and critic networks, the parameters of the target actor and critic networks are updated using 
\begin{equation}
    \theta_a' = \tau\theta_a + (1-\tau)\theta_a',
\end{equation}
and    
\begin{equation}
\theta_c' = \tau\theta_c + (1-\tau)\theta_c',
\end{equation}
respectively.

\begin{algorithm}[htbp]
\setstretch{1.35} 
\caption{DDPG Algorithm}
\label{DDPG}
\begin{algorithmic}
\REQUIRE 
Random process for exploration noise generation, $\mathcal{N}$;\\
Replay buffer size, $r$;\\
Size of minibatch for experience replay, $m$;\\
Coefficient for updating network parameters, $\tau$;\\
\STATE Initialize parameters of actor and critic networks, $\theta_a$ and $\theta_c$;\\
\STATE Initialize parameters of target actor and critic networks, $\theta_a' \leftarrow \theta_a$ and $\theta_c' \leftarrow \theta_c$;\\
\STATE Initialize the replay buffer, $\mathcal{R}$;\\
\FOR{$ep = 1$ to $EP_{max}$}
\STATE Reset the environment;\\
\FOR{$t = 1$ to $T$}
\STATE Observe the state of environment, $S(t)$;\\
\STATE Generate output of actor network, $O(t)$;\\
\STATE Invoke Alg. \ref{actionGeneration} to generate action $A(t)$;\\
\STATE Add noise for exploration, $A(t) = A(t) + \mathcal{N}(t)$;\\
\STATE Get immediate reward $f(t)$;\\
\STATE Observe the new state of environment, $S(t+1)$;\\
\STATE Store transition $(S(t), A(t), f(t), S(t+1))$ in $\mathcal{R}$;\\
\STATE Randomly select a minibatch of $m$ transitions from $\mathcal{R}$, $\{S_i, A_i, f_i, S_{i+1} | \; i \in \{1, \cdots, m\}\}$;\\
\STATE $\mathcal{L}_c = \frac{1}{m}\sum_{i = 1}^{m}(f_i + \gamma \chi(S_{i+1}, O(S_{i+1}, \theta_a'), \theta_c') - \chi(S_i, A_i, \theta_c))^2$;\\
\STATE Update the parameters of critic network, $\theta_c$, by minimizing $\mathcal{L}_c$;\\
\STATE $d\theta_a \leftarrow \Delta_{\theta_a}\frac{1}{m}\sum_{i = 1}^m\chi(S_i, O(S_i, \theta_a), \theta_c)$, $\theta_a \leftarrow \theta_a + \lambda(ep) \times d\theta_a$;\\
\STATE Update the parameters of the target actor and critic networks:\\
\STATE $\theta_a' = \tau\theta_a + (1-\tau)\theta_a'$, $\theta_c' = \tau\theta_c + (1-\tau)\theta_c'$;\\
\ENDFOR
\ENDFOR
\ENSURE The parameters of target actor and critic networks, $\theta_a'$ and $\theta_c'$;\\
\end{algorithmic}
\end{algorithm}

\section{Knowledge Embedding}
In this paper, we come up with a natural idea to improve the performance of DDPG-based packet schedulers by embedding human insights into them, called knowledge embedding. In knowledge embedding, the actions generated by trained DDPG agents are adjusted to meet the requirements of human insights in order to improve the performance of DDPG-based packet schedulers. Specifically, for the formulated problem of multi-user multi-application packet scheduling, we have some human insights as shown below.

\noindent \textbf{Human Insights:}

\emph{No PRB should be allocated to any application which is not running or any application of any UE with CQI value 0, in order to avoid waste of communication resource.}

According to the human insights, we have the knowledge embedding algorithm, as shown in Alg. \ref{KEDRL}. From Alg. \ref{KEDRL} we can find that if the CQI value of an UE is 0 or an application is not in use during time slot $t$, the corresponding value of the output of actor network of DDPG will be set to 0 if it is not 0. By doing this, it could be ensured that the generated action meets the requirement of human insights to avoid resource waste.

\begin{algorithm}[htbp]
\setstretch{1.35} 
\caption{Knowledge Embedding Algorithm}
\label{KEDRL}
\begin{algorithmic}
\REQUIRE The trained actor network of DDPG-based packet scheduler with parameters $\theta_a$;\\
\FOR{$t = 1$ to  $T$}
\STATE Observe the state of environment, $S(t)$;\\
\STATE Generate output of actor network, $O(t)$;\\
\FOR{$i = 1$ to $U$}
\FOR{$j = 1$ to $K$}
\IF{$I_i(t) == 0$ or $\sigma_i^j(t) == 0$}
\IF{$O_{(i-1) \times K + j}(t) \neq 0$}
\STATE $O_{(i-1) \times K + j}(t) = 0$;\\
\ENDIF
\ENDIF
\ENDFOR
\ENDFOR
\STATE Invoke Alg. \ref{actionGeneration} to generate action $A(t)$;\\
\ENDFOR
\end{algorithmic}
\end{algorithm}

\section{Simulation}
\subsection{Experimental Settings}
In order to mimic the real-world scenarios, we assume that the packet arrival of each application follows Poisson process. And it is assumed that during each episode, each application on each UE is randomly chosen to have packet arrival or not. In addition, an application is defined to be active only when its buffer is not empty. Moreover, for better imitating the scenes in the actual world, it is assumed that at most one application in UHD video streaming, web browsing and online gaming can be active within each episode on each UE. The parameter settings of the five selected applications are summarized in Table \ref{Parameter_settings}.

\begin{table*}[htbp]

\centering

\caption{Parameter Settings of the Five Selected Applications.}

\label{Parameter_settings}

\begin{tabular}{c|c|c|c|c|c}

  \hline
      \diagbox[width = 140pt]{Parameter}{Application} & \textrm{FTP} & \textrm{UHD video streaming} & \textrm{Web browsing} & \textrm{Online gaming} & \textrm{VoIP}\\
  \hline
  Data rate requirement & 0 & 3 Mbps & 0 & 5 Mbps & 60 Kbps\\
  \hline
  Latency requirement & 300 ms & 300 ms & 300 ms & 50 ms & 100 ms\\
  \hline
  Packet loss rate requirement & 0.5\% & 1\% & 1\% & 1\% & 1\%\\
  \hline
  Packet size & 1500 Bytes & 1500 Bytes & 150 Bytes & 150 Bytes & 20 Bytes\\
  \hline
  Mean packet arrival rate& 500 pps & 300 pps & 1000 pps & 5000 pps & 500 pps\\
  \hline
  Packet delay budget & 300 ms & 300 ms & 300 ms & 50 ms & 100 ms\\
  \hline
  Priority & 1 & 2 & 1 & 3 & 4\\
  \hline
\end{tabular}

\end{table*}

The main experimental settings are summarized in Table \ref{Main_parameter_settings}. As mentioned above, we don't have the knowledge of the transition probability distribution of the problem of multi-user multi-application packet scheduling since that the environment changes dynamically and randomly. However, in order to conduct simulation, for mimicking the real-world scenarios, for each episode, the CQI values of UEs are randomly initialized according to the probabilities shown in Table \ref{CQI_probabilities}. Moreover, an assumption is made that the CQI value of each UE changes every 40 time slots. The probability of the CQI value of an UE to be the same is set to be 80\%. The probabilities of the CQI value of an UE to be 1 higher and 1 lower are set to be both 10\%.

\begin{table}[htbp]

\centering

\caption{Main Experimental Settings.}

\label{Main_parameter_settings}

\begin{tabular}{c|c}

  \hline
  \textrm{Parameter} & \textrm{Value}\\
  \hline
  Number of UEs & 10\\
  \hline
  Number of applications & 5\\
  \hline
  Average latency from server to base station & 22 ms\\
  \hline
  Length of one TTI & 1 ms\\
  \hline
  Length of time window for calculating & \multirow{2}{*}{500 ms}\\
  packet loss rate and burst ratio & \\
  \hline
  Discount factor & 0.99\\
  \hline
  Length of one episode & 1 minute\\
  \hline
  Number of time slots per episode & 60000\\
  \hline
  Number of training episodes & 200\\
  \hline
  Number of testing episodes & 100\\
  \hline
  Buffer size of DDPG algorithm & 50000\\
  \hline
  Minibatch size for experience replay & \multirow{2}{*}{64}\\
  in DDPG algorithm & \\
  \hline
  Coefficient for updating network & \multirow{2}{*}{0.005}\\
  parameters in DDPG algorithm & \\
  \hline
\end{tabular}

\end{table}

\begin{table}[htbp]

\centering

\caption{CQI Value Initialization Probabilities.}

\label{CQI_probabilities}

\begin{tabular}{c|c}

  \hline
  \textrm{Initial CQI value} & \textrm{Probability}\\
  \hline
  0, 1, 2, 3& 1\% for each value\\
  \hline
  4, 5, 6 & 2\% for each value\\
  \hline
  7, 8, 9, 10, 11, 12, 13, 14, 15 & 10\% for each value\\
  \hline
\end{tabular}

\end{table}

\begin{table}[htbp]

\centering

\caption{CQI-MCS Mapping.}

\label{CQI_MCS_Mapping}

\begin{tabular}{c|c|c|c}

  \hline
  \textrm{CQI} & \textrm{Modulation} & \textrm{Code rate} & \textrm{Useful bits per}\\ 
  \textrm{Value} & \textrm{Scheme} & \textrm{$\times$ 1024} & \textrm{symbol carries}\\
  \hline
  0 & \textrm{None} &  \textrm{None} &  \textrm{None}\\
  \hline
  1 & \textrm{QPSK} & 78 & 0.1523\\
  \hline
  2 & \textrm{QPSK} & 193 & 0.3770\\
  \hline
  3 & \textrm{QPSK} & 449 & 0.8770\\
  \hline
  4 & \textrm{16QAM} & 378 & 1.4766\\
  \hline
  5 & \textrm{16QAM} & 490 & 1.9141\\
  \hline
  6 & \textrm{16QAM} & 616 & 2.4063\\
  \hline
  7 & \textrm{64QAM} & 466 & 2.7305\\
  \hline
  8 & \textrm{64QAM} & 567 & 3.3223\\
  \hline
  9 & \textrm{64QAM} & 666 & 3.9023\\
  \hline
  10 & \textrm{64QAM} & 772 & 4.5234\\
  \hline
  11 & \textrm{64QAM} & 873 & 5.1152\\
  \hline
  12 & \textrm{256QAM} & 711 & 5.5547\\
  \hline
  13 & \textrm{256QAM} & 797 & 6.2266\\
  \hline
  14 & \textrm{256QAM} & 885 & 6.9141\\
  \hline
  15 & \textrm{256QAM} & 948 & 7.4063\\
  \hline
\end{tabular}

\end{table}

\begin{table}[htbp]

\centering

\caption{Learning Rate Scheduling of DDPG Algorithm.}

\label{learning_rate_scheduling}

\begin{tabular}{c|c}

  \hline
  \textrm{Episode} & \textrm{Learning rate}\\
  \hline
  1 - 50 & 0.00001\\
  \hline
  51 - 100 & 0.000001\\
  \hline
  101 - 150 & 0.0000001\\
  \hline
  151 - 200 & 0.00000001\\
  \hline
\end{tabular}

\end{table}

\begin{table}[htbp]

\centering

\caption{Standard Deviation Scheduling for Exploration Noise Generation in DDPG Algorithm.}

\label{STD_scheduling}

\begin{tabular}{c|c}

  \hline
  \textrm{Episode} & \textrm{Standard deviation of added noise}\\
  \hline
  1 - 50 & 0.02\\
  \hline
  51 - 100 & 0.002\\
  \hline
  101 - 200 & 0\\
  \hline
\end{tabular}

\end{table}

For modelling the process of multi-user multi-application packet scheduling in downlink 6G RAN realistically, in this paper, the CQI-MCS mapping introduced in \cite{3GPP26213} is employed, shown as Table \ref{CQI_MCS_Mapping}. The learning rate scheduling for the training of DDPG agents is shown in Table \ref{learning_rate_scheduling}. For exploration noise generation in DDPG algorithm, Gaussian distribution $\mathcal{N}(0, \omega(t)^2)$ is utilized to generate the noise value added to each element of the action generated at time slot $t$, where $\omega(t)$ denotes the standard deviation of Gaussian distribution. The scheduling of standard deviation for exploration noise generation in DDPG algorithm is shown in Table \ref{STD_scheduling}.

Moreover, in order to investigate the influences of the settings of number of middle layers of actor and critic networks, represented by $p$ and $q$ in Fig. \ref{actorNetwork} and \ref{criticNetwork-DDPG}, on the performance of the DDPG-based packet schedulers, four different settings are selected. In setting 1, $p = 2$, $q = 1$. In setting 2, $p = 3$, $q = 2$. In setting 3, $p = 4$, $q = 3$. In setting 4, $p = 5$, $q = 4$. 

\begin{figure*}[bp]
  \begin{center}
  \includegraphics[width=7in]{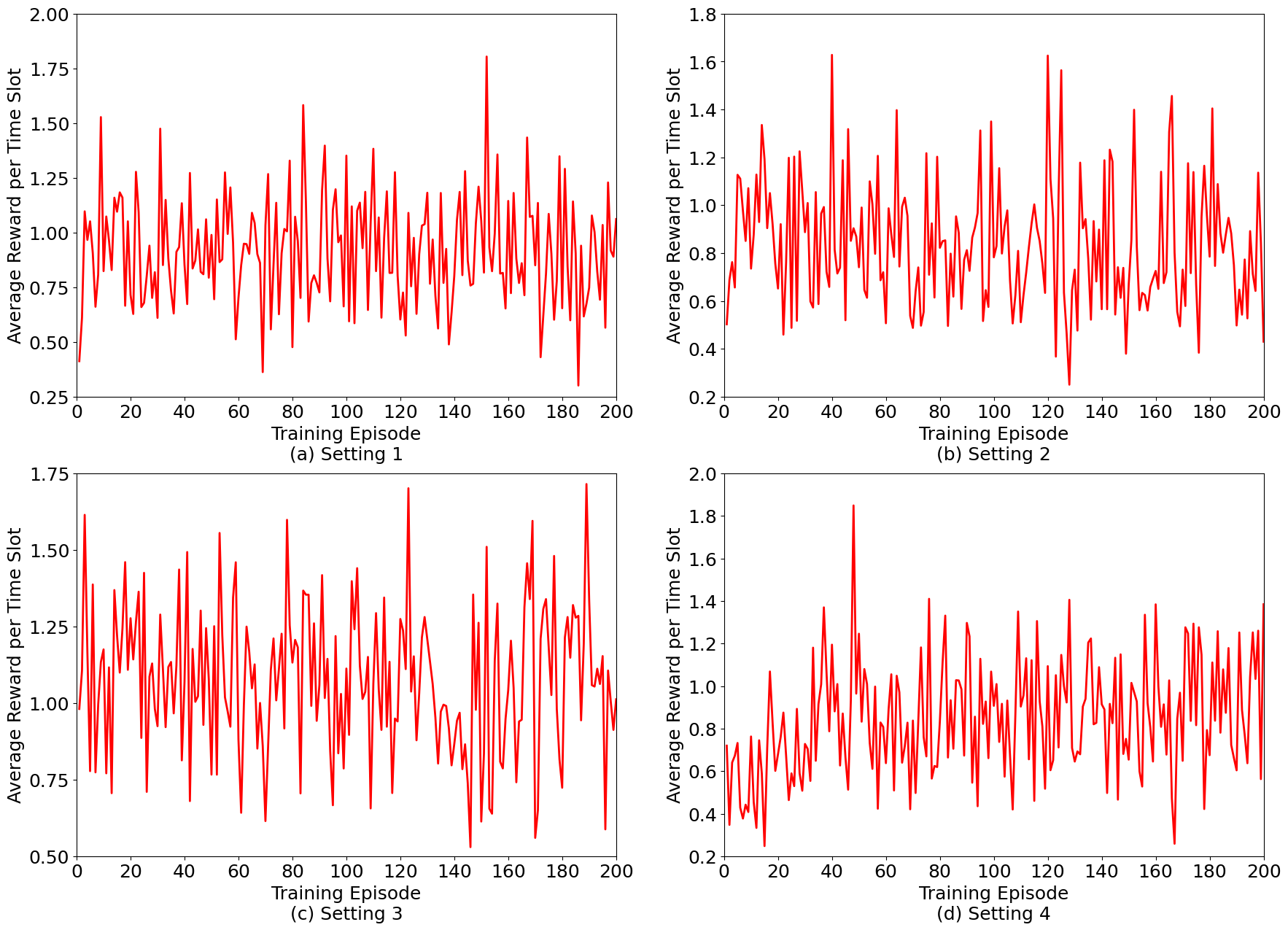}\\
  \caption{Simulation results of DDPG algorithm under the four settings during the training stage.}\label{TrainPerformance}
  \end{center}
\end{figure*}

\begin{figure*}
  \begin{center}
  \includegraphics[width=7in]{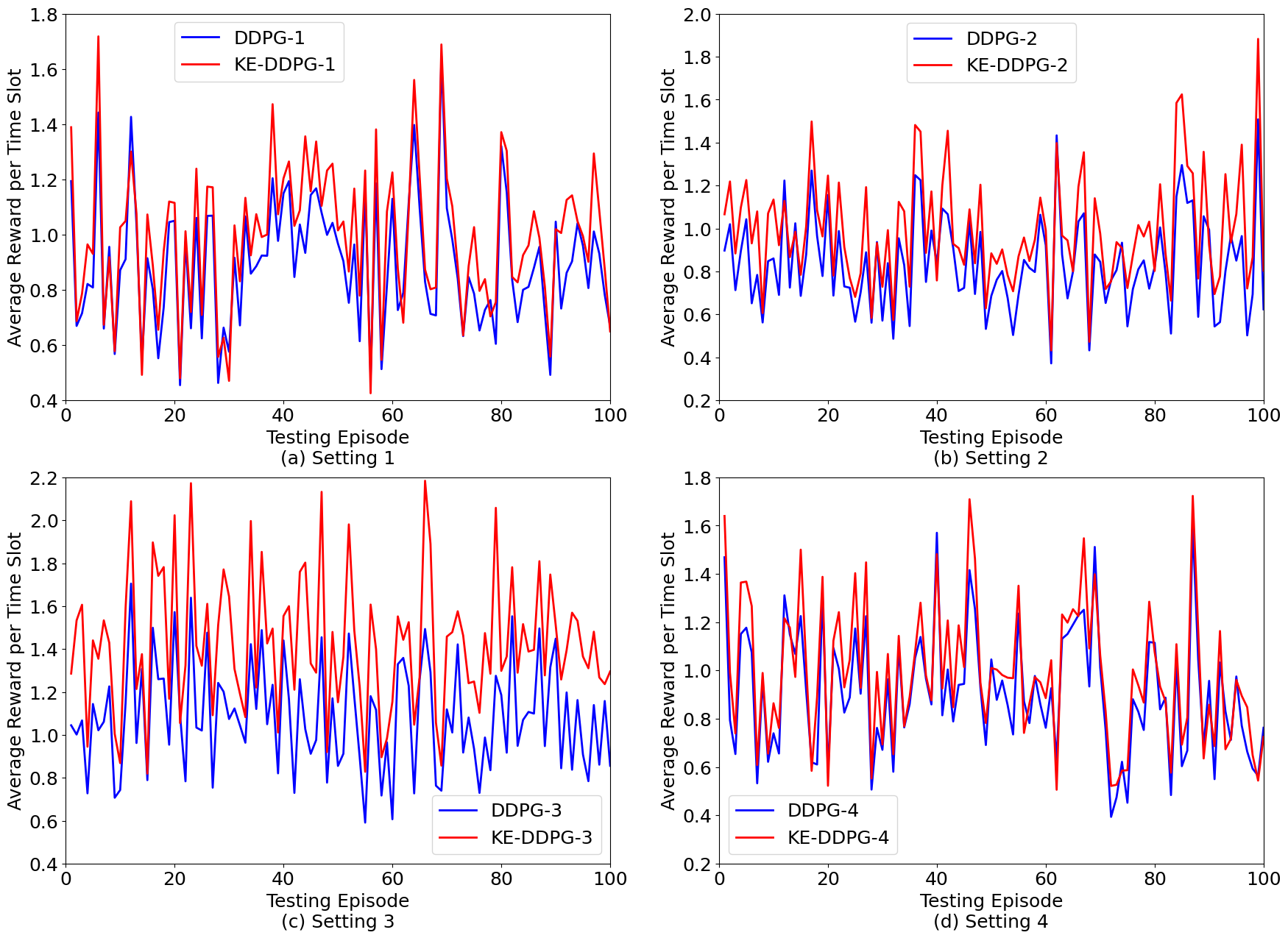}\\
  \caption{Performance comparison of DDPG and KE-DDPG-based packet schedulers under the four settings on the test dataset. Under each setting, under the majority of testing episodes, KE-DDPG-based packet scheduler outperforms DDPG-based packet scheduler, indicating that knowledge embedding is effective to improve the performance of DDPG-based packet schedulers.}\label{TestPerformance-1}
  \end{center}
\end{figure*}

\subsection{Simulation Results During the Training Stage}
The simulation results of DDPG algorithm under the four settings during the training stage are shown in Fig. \ref{TrainPerformance}. From Fig. \ref{TrainPerformance}, we can find that the average reward per time slot fluctuates over the training episode. One core reason is that the random initialization of CQI values of UEs greatly influences the achieved rewards during each episode. In an episode where the initial CQI values of UEs are high, it could be easier to get a higher sum of rewards during this episode. However, in an episode where the initial CQI values of UEs are low, it could be more possible to get a lower sum of rewards during this episode, since that it's more difficult to satisfy the QoS requirements of applications. In addition, the random initialization of CQI values of UEs inevitably makes it more difficult for the DDPG agents to successfully apply the experience they gained from previous training episodes during the current training episode, as the gained experience could be strongly relied on the corresponding CQI values.

\subsection{Comparison Algorithms}
 Five classic algorithms are selected as comparison algorithms, which are round robin (RB) \cite{6226795}, maximum throughput (MT) \cite{4556220}, proportional fair (PF) \cite{5431383}, earliest deadline first (EDF) \cite{6488471} and largest weighted delay first (LWDF) \cite{singh2013radio}.

\subsection{Testing Performance}
To better conduct performance comparison, a test dataset consisted of 100 episodes is created. The performance comparison of DDPG and knowledge embedded-DDPG (KE-DDPG)-based packet schedulers under the four settings of numbers of middle layers in actor and critic networks is shown in Fig. \ref{TestPerformance-1}. From Fig. \ref{TestPerformance-1}, we can find that under all the four settings, knowledge embedding could improve the performance of DDPG-based packet schedulers under most episodes. These results confirm the effectiveness of knowledge embedding.

\begin{figure}
  \begin{center}
  \includegraphics[width=3.5in]{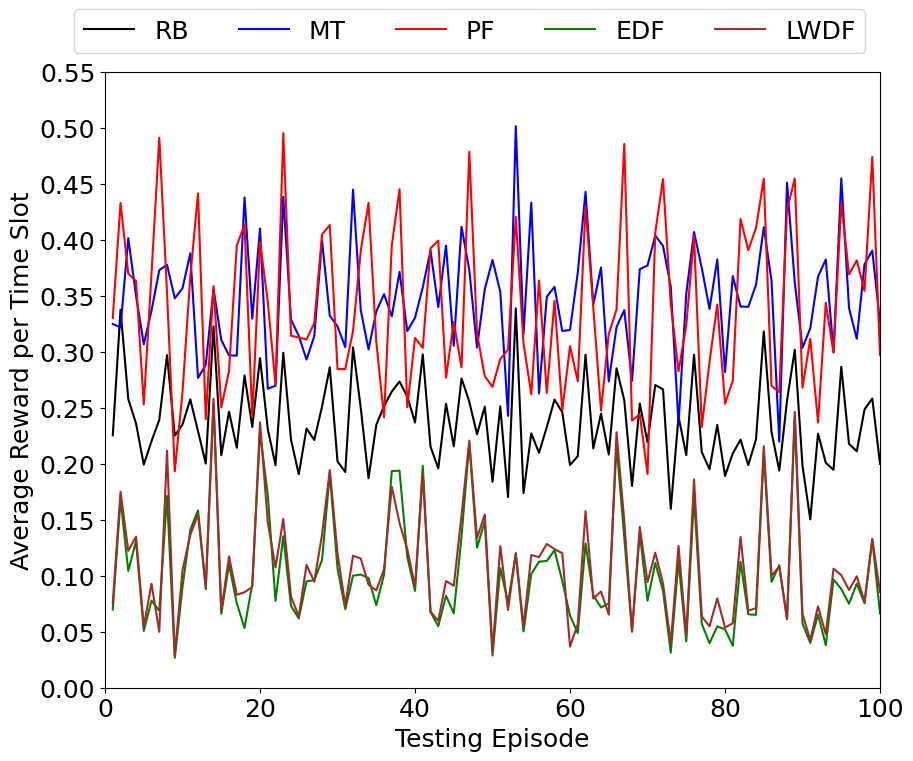}\\
  \caption{Performance comparison of RB, MT, PF, EDF and LWDF algorithms on the test dataset.}\label{TestPerformance-2}
  \end{center}
\end{figure}

The performance comparison of the five baseline algorithms is shown in Fig. \ref{TestPerformance-2}. From Fig. \ref{TestPerformance-2}, we can find that generally speaking, MT and PF achieve the top 2 performance on the test dataset. In addition, EDF and LWDF achieve the worst performance on the test dataset.

\begin{figure}
  \begin{center}
  \includegraphics[width=3.5in]{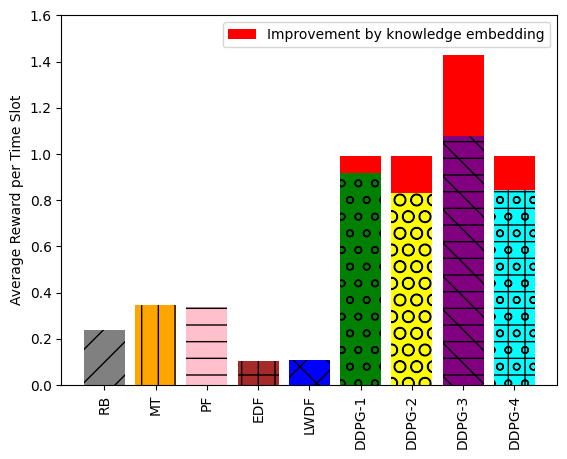}\\
  \caption{Comparison of average reward per time slot among DDPG-based packet schedulers and baseline algorithms on the test dataset.}\label{Average_reward}
  \end{center}
\end{figure}

The comparison of average reward per time slot on the whole test dataset is shown in Fig. \ref{Average_reward}. From Fig. \ref{Average_reward}, we can find that under each setting, DDPG-based packet scheduler achieves much better performance than the baseline algorithms. For example, the average reward per time slot of RB, MT, PF, EDF and LWDF are only 21.94\%, 32.31\%, 31.29\%, 9.46\% and 10.12\% of that of DDPG-based packet scheduler under setting 3, respectively. These results demonstrate the superiority of DDPG-based packet scheduler over the baseline algorithms. In addition, we could notice that under each setting, the performance of DDPG-based packet scheduler is largely improved after using knowledge embedding. By using knowledge embedding, the performance of the DDPG-based packet schedulers under setting 1-4 are improved by 8.07\%, 18.92\%, 32.59\% and 17.20\%, respectively. These results confirm the effectiveness of knowledge embedding on improving the performance of DDPG-based packet schedulers.

\section{Conclusion}

In this paper, in order to achieve application-specific QoE enhancement, the problem of multi-user multi-application packet scheduling in downlink 6G RAN is investigated. For measuring fairnesses among UEs and their active applications, intra-UE fairness and inter-UE fairness are proposed. This problem is formulated as a sequential decision-making problem with the objective of maximizing the sum of inter-UE fairnesses of a time period. This problem is further reformulated as a Markov decision process (MDP) problem. For solving this MDP problem, a DDPG-based solution is designed. Due to the high dimensionalities of both the state and action spaces, it's difficult for the DDPG agents to grasp some human insights through exploration and exploitation by themselves. Therefore, a novel knowledge embedding method is proposed to avoid resource wastes and improve the performance of the trained DDPG agents. In order to realistically conduct performance evaluation, five representative applications are considered. And to fairly compare the performance of algorithms, a test dataset is created. A plethora of experiments are conducted, whose results confirm the superiority of DDPG-based packet schedulers over the baseline algorithms. In addition, the simulation results also confirm the effectiveness of knowledge embedding on improving the performance of DDPG-based packet schedulers.

\bibliographystyle{IEEEtran}
\bibliography{IEEEabrv,Bibliography}

\begin{IEEEbiography}[{\includegraphics[width=1in,height=1.25in,clip,keepaspectratio]{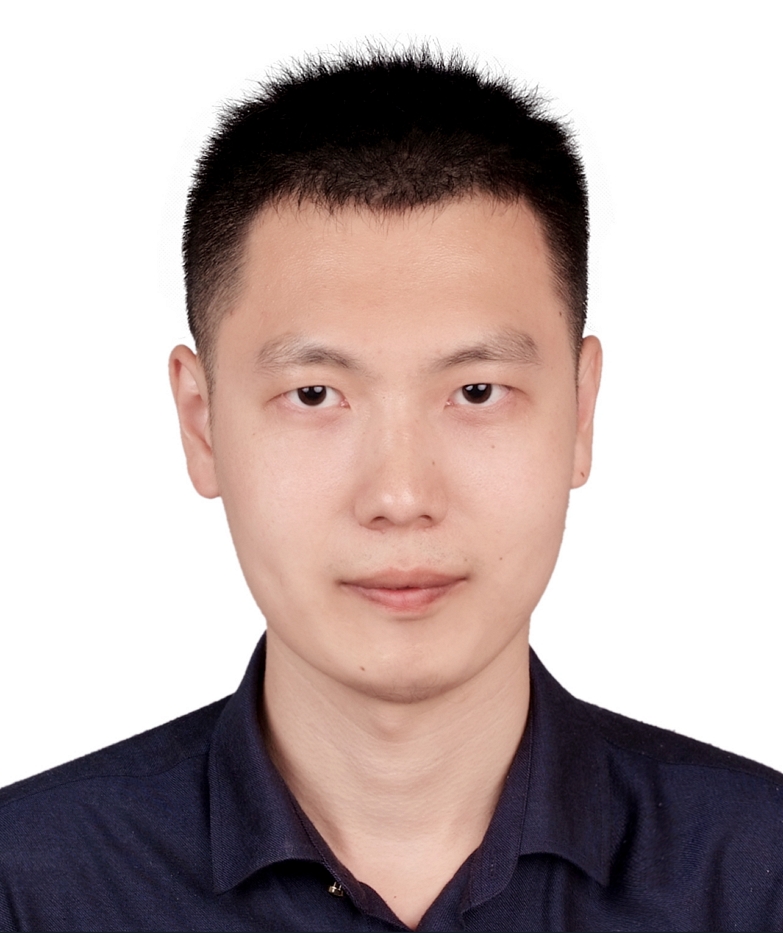}}]{Yongqin Fu}
    (Graduate Student Member, IEEE) received the B.E. degree in information engineering from Southeast University, Nanjing, China, in 2015 and the M.E. degree in computer science and technology from Zhejiang University, Hangzhou, China, in 2019. He is currently working towards the Ph.D. degree at the Department of Electrical and Computer Engineering, Western University, London, Ontario, Canada. 
    
    His current research interests include intelligent and customized resource management and network cooperation in beyond 5G and 6G networks, as well as sensor data analysis in Internet of Things (IoT) systems.
\end{IEEEbiography}

\begin{IEEEbiography}[{\includegraphics[width=1in,height=1.25in,clip,keepaspectratio]{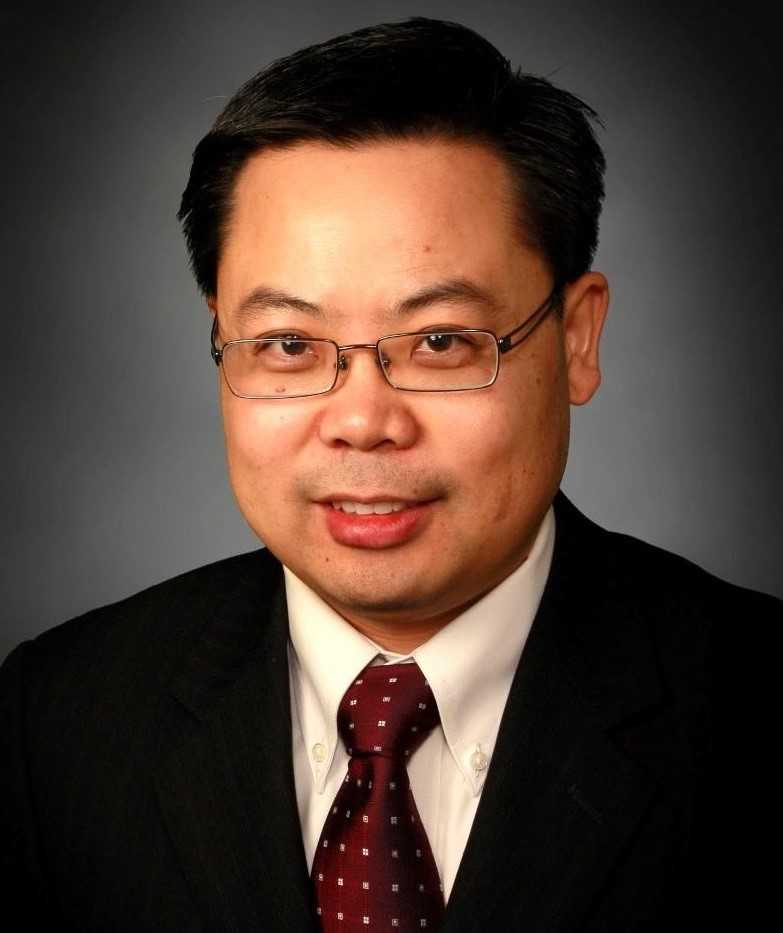}}]{Xianbin Wang}
   (Fellow, IEEE) received his Ph.D. degree in electrical and computer engineering from the National University of Singapore in 2001.
   
He is a Professor and a Tier-1 Canada Research Chair in 5G and Wireless IoT Communications with Western University, Canada. Prior to joining Western University, he was with the Communications Research Centre Canada as a Research Scientist/Senior Research Scientist from 2002 to 2007. From 2001 to 2002, he was a System Designer at STMicroelectronics. His current research interests include 5G/6G technologies, Internet of Things, communications security, machine learning, and intelligent communications. He has over 500 highly cited journals and conference papers, in addition to over 30 granted and pending patents and several standard contributions.

Dr. Wang is a Fellow of the Canadian Academy of Engineering and a Fellow of the Engineering Institute of Canada. He has received many prestigious awards and recognitions, including the IEEE Canada R. A. Fessenden Award, Canada Research Chair, Engineering Research Excellence Award at Western University, Canadian Federal Government Public Service Award, Ontario Early Researcher Award, and nine Best Paper Awards. He was involved in many IEEE conferences, including GLOBECOM, ICC, VTC, PIMRC, WCNC, CCECE, and CWIT, in different roles, such as General Chair, TPC Chair, Symposium Chair, Tutorial Instructor, Track Chair, Session Chair, and Keynote Speaker. He serves/has served as the Editor-in-Chief, Associate Editor-in-Chief, and editor/associate editor for over ten journals. He was the Chair of the IEEE ComSoc Signal Processing and Computing for Communications (SPCC) Technical Committee and is currently serving as the Central Area Chair for IEEE Canada.

\end{IEEEbiography}

\vfill

\end{document}